\documentclass[10pt,a4paper]{article}   

\usepackage{amsmath}    
\usepackage{amsfonts}  
\usepackage{amssymb}
\usepackage{graphicx}   
\usepackage{soul} 

\usepackage[utf8]{inputenc}
\usepackage[T1]{fontenc}

\usepackage{float}   

\usepackage[a4paper]{geometry} 
\usepackage{color}   
\usepackage{soulutf8}    
\definecolor{lightblue}{rgb}{0.8,0.9,1} 
\definecolor{lightred}{rgb}{1,0.5,0.4} 
\definecolor{lightgreen}{rgb}{0.4,1,0.4} 

\newtheorem{Theoreme}{THEOREM}
\newtheorem{Definition}{Definition}
\newtheorem{Hypothese}{Hypothesis}

\newtheorem{Proposition}{Proposition}
\newtheorem{Remarque}{Remark}

\begin{document}

\title{ Double-scale theory }
\author{Michel Gondran\\\textit{Académie Européenne Interdisciplinaire des Sciences, Paris, France}\\\texttt{michel.gondran@polytechnique.org}\\~\\
Alexandre Gondran\\\textit{\'Ecole Nationale de l'Aviation Civile, Toulouse,France}\\\texttt{alexandre.gondran@recherche.enac.fr}}

\date{January 2023}
\maketitle

\begin{abstract}


We present a new interpretation of quantum mechanics, called the double-scale theory, which expends on the de Broglie-Bohm  (dBB) theory. It is based, for any quantum system, on the simultaneous existence of two wave functions in the laboratory reference frame~: an external wave function and an internal one. The external wave function corresponds to a field that pilots the center-of-mass of the quantum system. The external wave spreads out in space over time. Mathematically, the Schrödinger equation converges to the Hamilton-Jacobi statistical equations when the Planck constant tends towards zero and the Newton trajectories are therefore approximations of the dBB trajectories.

The internal wave function 
corresponds to the interpretation proposed by Edwin Schrödinger for whom the particle is extended. Then, the internal wave remains confined in space. Its converges, when $\hbar\rightarrow 0$, to a Dirac distribution.
Furthermore, we show that non-stationary solutions can exist such that the $3N$-dimensional configuration space of the internal wave function can be rewritten as the product of $N$ individual 3-dimensional internal wave functions.

\end{abstract}

\maketitle


\section{Introduction}

Following de Broglie's idea, we presented recently~\cite{Gondran2021} a new interpretation of quantum mechanics that defined two waves functions for any quantum system~: an external one and an internal one. 
In this paper, we clarify this interpretation, which we name the double-scale theory, and in particular we demonstrate that the two wave functions exist simultaneously in the same reference frame, that of the laboratory.

These two functions have very different physical behaviors~: the external wave function spreads out over time contrary to the internal wave function which remains confined. To justify the distinction between these two wave functions, we demonstrate that the mathematical convergences are fundamentally different when $\hbar\rightarrow0$. For the external wave function, its phase and the square of its modulus converge towards an action and a classical density satisfying the Hamilton-Jacobi statistical equations. For the internal wave function, the square of its modulus converges to a Dirac distribution, thus losing the internal structure of the atom or molecule.

We deduce very different interpretations for these two wave functions. By mathematical continuity, we  show that the phase of the external wave \textit{``drives''} the center-of-mass of the quantum system and corresponds to the de Broglie-Bohm interpretation restricted to the external wave function. The most likely interpretation of the internal wave is the one proposed by Schrödinger for which the particles are extended.

The plan of the paper is as follows. In section~\ref{section:centerOfMass}, we recall the well-known existence of the two wave functions: the center-of-mass wave function in the laboratory reference frame and the relative wave function in the center-of-mass reference frame.

 
In section~\ref{section:externalWave}, we show that the center-of-mass wave function in the laboratory reference frame is an external wave function that corresponds to a non-local field that \textit{``pilots''} the center-of-mass of the corpuscle and that converges to classical mechanics when we make Planck's constant tends to zero. This is the de Broglie-Bohm interpretation restricted to the external wave function.

In section~\ref{section:InternalWave}, from the value of the center-of-mass and of the relative  wave function, we define an internal wave function in the laboratory reference frame, that is local and corresponds to the corpuscle in the field-corpuscle duality. This is the Schrödinger interpretation restricted to the internal wave function. Moreover, we show that non-stationary solutions can exist such that the $3N$-dimensional configuration space of the internal wave function can be rewritten as the product of $N$ individual internal wave functions in the 3 dimensional space.

In section~\ref{section:NewReading}, the simultaneous existence of these two functions allows us to understand and explain simply many phenomena of quantum mechanics, such as wave-particle duality, quantum measurements and the nonlocality of the EPR-B experiment. 

\section{Center-of-mass wave and relative wave}
\label{section:centerOfMass} 


Since the beginning of quantum mechanics, two types of wave functions have been distinguished in order to study the dynamics of atoms and molecules.
Indeed, it is well known to decompose the total wave function into the \emph{center-of-mass wave function} in the laboratory reference frame and the \emph{relative wave function} in the center-of-mass reference frame.

Roughly, the center-of-mass wave function, renamed \textbf{external wave function}, corresponds to the external dynamics of the atom, i.e. the motion of its center-of-mass and the orientation of the reference frame that is linked to it. The relative wave function describes the structure of the atom or molecule in the reference frame of its center-of-mass.
By rewriting the relative wave function in the laboratory reference frame, we obtain the so-called \textbf{internal wave function}. The double-scale theory is built on the existence of these two wave functions (external and internal) defined in the laboratory reference frame and by their interpretation. In this section we recall this decomposition for a quantum system with $N$ bodies (atoms or molecules) where the decomposition is exact.

Let us consider a system of $N$ spinless particles of masses $m_j$ and charges $q_j$ (with $j=1..N$), of coordinates $\mathbf{x}_j$, subjected to an 
external gravitational field $V_g(\mathbf{x}_j)$ and a zero external electric field, and to mutual interactions described by the potentials $U_{jk}(\vert\mathbf{x}_j -\mathbf{x}_k\vert),\forall j,k=1...N,\ j\neq k$. 
This quantum system is then described by the wave function $ \Psi(\mathbf{x}_1, \mathbf{x}_2,...,\mathbf{x}_N, t) $ which verifies the Schrödinger equation:
\begin{equation}\label{eq:schrodinger1d}
i\hslash \frac{\partial \Psi(\mathbf{x}_1, \mathbf{x}_2,.., \mathbf{x}_N, t) }{\partial t}= H \Psi(\mathbf{x}_1, \mathbf{x}_2,..,\mathbf{x}_N, t) 
\end{equation}
with the Hamiltonian:
\begin{equation}\label{eq:hamiltonschrodinger1}
H= \sum_{j=1}^N \left(\frac{\textbf{p}_j ^{2}}{2m_j} + m_j V_g(\mathbf{x}_j)\right)+ \sum_{j=1}^N\sum_{\underset{k \neq j}{k=1}}^N U_{jk}(\vert \mathbf{x}_j -\mathbf{x}_k \vert)
\end{equation}
and the initial condition~: 
\begin{equation}\label{eq:schrodinger2}
\Psi (\mathbf{x}_1, \mathbf{x}_2,.., \mathbf{x}_N, 0)=\Psi_{0}(\mathbf{x}_1,\mathbf{x}_2 , ..,\mathbf{x}_N).
\end{equation}

We separate the motion of these $N$ particles from the motion of their center-of-mass: let $\mathbf{x}_G=(\sum_{j=1}^N m_j \mathbf{x}_j)/(\sum_{j=1}^N m_j)$ be the position of the center-of-mass, $\mathbf{x}_j'= \mathbf{x}_j -\mathbf{x}_G$ be the relative position of the $j$ particle and $M= \sum_{j=1}^N m_j$ be the total mass.

 Then the Hamiltonian $H$ is written as a function of the total impulse ($\textbf{p}_G= \sum_{i=1}^N \textbf{p}_i $) and relative impulses ($\textbf{p}'_i= \textbf{p}_i -m_i/M \textbf{p}_G $) taking into account small variations of the gravitational field
 $V_g(\mathbf{x}_j)\simeq V_g(\mathbf{x}_G) + \mathbf{x}'_j\nabla V_g(\mathbf{x}_G)$:
\begin{equation}\label{eq:hamiltonschrodinger2}
H= \left( \frac{\textbf{p}_G^{2}}{2M} + M V_g(\mathbf{x}_G) \right)+ \sum_{i=1}^N \left(\frac{\textbf{p}_i'^{2}}{2 m_i} + \sum_{\underset{j \neq i}{j=1}}^N U_{ij}(\vert \mathbf{x}'_i -\mathbf{x}'_j\vert )\right) = H^{ext} + \sum_{i=1}^NH_i^{int}. 
\end{equation}
We note that the interaction of the gravitational field related to its local variation does not intervene because $\sum_{i=1}^N m_i (\mathbf{x}_i- \mathbf{x}_G)=0$.

\begin{Proposition}\label{prop:3} - If the initial wave function $ \Psi_{0}(\mathbf{x}_1,\mathbf{x}_2,..,\mathbf{x}_N)$ factors to the form:  
\begin{equation}\label{eq:solschrodinger2}
 \Psi_0(\mathbf{x}_1, \mathbf{x}_2,.., \mathbf{x}_N) =\psi_0(\mathbf{x}_G) \varphi_0(\mathbf{x}'_1,\mathbf{x}'_2,..,\mathbf{x}'_N).
\end{equation}
then $ \Psi(\mathbf{x}_1, \mathbf{x}_2,.., \mathbf{x}_N, t)$, 
a solution to (\ref{eq:schrodinger1d}),(\ref{eq:hamiltonschrodinger1}) and (\ref{eq:schrodinger2}), is written as the product of the center-of-mass wave function $ \psi(\mathbf{x}_G,t)  $ and the relative function $\varphi(\mathbf{x}'_1,\mathbf{x}'_2,..,\mathbf{x}'_N,t)$~:
\begin{equation}\label{eq:solschrodinger3b}
 \Psi(\mathbf{x}_1, \mathbf{x}_2,.., \mathbf{x}_N,t) =\psi(\mathbf{x}_G,t) \varphi(\mathbf{x}'_1,\mathbf{x}'_2,..,\mathbf{x}'_N, t)
\end{equation}
where $\psi(\mathbf{x}_G,t) $ is the solution to the Schrödinger equations of the center-of-mass:
\begin{equation}\label{eq:schrodinger1}
i\hslash \frac{\partial \psi(\mathbf{x}_G,t) }{\partial t}=- \frac{\hbar ^{2}}{2M}
\Delta_{\mathbf{x}_G} \psi(\mathbf{x}_G,t)+ M V_g(\mathbf{x}_G) \psi(\mathbf{x}_G,t)
\end{equation}
with the initial condition:
\begin{equation}\label{eq:solschrodinger4}
\psi(\mathbf{x}_G,0) =\psi_0(\mathbf{x}_G)
\end{equation}
and where $ \varphi(\mathbf{x}'_1,\mathbf{x}'_2,..,\mathbf{x}'_N,t)$ is the Schrödinger relative equations:
\begin{equation}\label{eq:schrodinger2b}
i\hslash \frac{\partial \varphi(\mathbf{x}'_1,\mathbf{x}'_2,..,\mathbf{x}'_N,t) }{\partial t}= - \sum_{i=1}^N \left(\frac{\hbar ^{2}}{2 m_i}
\Delta_{\mathbf{x}'_i} + \sum_{\underset{j \neq i}{j=1}}^N U_{ij}(\vert\mathbf{x}'_i -\mathbf{x}'_j \vert)\right) \varphi(\mathbf{x}'_1,\mathbf{x}'_2,..,\mathbf{x}'_N,t)
\end{equation}
with the initial condition:
\begin{equation}\label{eq:solschrodinger6}
 \varphi(\mathbf{x}'_1,\mathbf{x}'_2,..,\mathbf{x}'_N,0)=  \varphi_0(\mathbf{x}'_1,\mathbf{x}'_2,..,\mathbf{x}'_N).
\end{equation}
\end{Proposition}

We have considered the case of a decomposition of the total wave function as the product of the center-of-mass wave function in the laboratory reference frame and the relative wave function in the center-of-mass reference frame. This example highlights the simultaneous existence of these two functions.
We postulate that the decomposition of the total wave function of an $N$-body quantum system $\Psi(\mathbf{x}_1, \mathbf{x}_2,.., \mathbf{x}_N,t)$ into two waves functions $\psi(\mathbf{x}_G,t)$ and $\varphi(\mathbf{x}'_1,\mathbf{x}'_2,..,\mathbf{x}'_N, t)$ is not only a mathematical tool to simplify the calculations, but that these two waves physically exist at all times and simultaneously.


When the quantum system is not composed of several particles but corresponds to a single particle, like a free electron, we postulate that these two wave functions exist simultaneously, even if the mathematical decomposition is no longer necessary.

\begin{Hypothese}- \textbf{Hypothesis for a single particle}: A single particle, like a free electron is also simultaneously described by two wave functions: an external wave function $\psi(\mathbf{x}_G,t)$ related to its center-of-mass and a relative wave function $\varphi (\mathbf{x}',t)$. This hypothesis assumes that the particle is not point-like and has an extension represented by the relative wave function; this is the Schrödinger conjecture of a single particle, witch we develop in section~\ref{section:InternalWave}.
\end{Hypothese}


 The center-of-mass wave function and the relative wave function are defined in different reference frames. We will call the center-of-mass wave function the external wave function and the relative wave function the internal wave function when it is rewritten (see section~\ref{section:InternalWave}) to be considered in the laboratory reference frame as the external wave function. These two wave functions, external and internal, have very different behaviors: the external wave function spreads over time and is non-local, while the internal wave function remains confined and is local. They will give rise to very different interpretations as we will see in the next two sections.
 
\section{The external wave function: the field of the field-corpuscle duality}
\label{section:externalWave}

To interpret the external wave function, let us study its convergence to classical mechanics when we make Planck's constant tends towards zero.

\subsection{Convergence of the external wave function}

Let us consider an external wave function verifying the external Schrödinger equations  
(\ref{eq:schrodinger1}) and (\ref{eq:solschrodinger4}) and make the semi-classical change of variable $\psi(\mathbf{x}_G,t)=\sqrt{\rho^{\hbar}(\mathbf{x}_G,t)} \exp\left(i
\frac{S^{\hbar}(\mathbf{x}_G,t)}{\hbar}\right)$. The density
$\rho^{\hbar}(\mathbf{x}_G,t)$ and the action $S^{\hbar}(\mathbf{x}_G,t)$ then verify the 
Madelung equations~\cite{Madelung1926} (1926):

\begin{equation}\label{eq:Madelung1}
\frac{\partial S^{\hbar}(\mathbf{x}_G,t)}{\partial t}+\frac{(\nabla S^{\hbar}(\mathbf{x}_G,t))^2}{2M} +
V(\mathbf{x}_G)-\frac{\hbar^2}{2M}\frac{\Delta
\sqrt{\rho^{\hbar}(\mathbf{x}_G,t)}}{\sqrt{\rho^{\hbar}(\mathbf{x}_G,t)}}=0
 \qquad \forall (\mathbf{x}_G,t)\in \mathbb{R}%
^{3}\times \mathbb{R}^{+}
\end{equation}

\begin{equation}\label{eq:Madelung2}
\frac{\partial \rho^{\hbar}(\mathbf{x}_G,t)}{\partial t}+
div\left(\rho^{\hbar}(\mathbf{x}_G,t)
\frac{\nabla S^{\hbar}(\mathbf{x}_G,t)}{m}\right)=0  \qquad \forall (\mathbf{x}_G,t)\in \mathbb{R}%
^{3}\times \mathbb{R}^{+}
\end{equation} 
with the initial conditions 
\begin{equation}\label{eq:Madelung3}
\rho^{\hbar}(\mathbf{x}_G,0)=\rho^{\hbar}_{0}(\mathbf{x}_G) \qquad \text{and}
\qquad S^{\hbar}(\mathbf{x}_G,0)=S^{\hbar}_{0}(\mathbf{x}_G) \qquad
\forall \mathbf{x}_G\in \mathbb{R}^{3}
\end{equation}

Here, $V(\mathbf{x}_G)=M V_g(\mathbf{x}_G)$ is the potential of (\ref{eq:schrodinger1}). 
Let us now study the convergence of
the density $\rho^{\hbar}(\mathbf{x}_G,t)$ and the action
$ S^{\hbar}(\mathbf{x}_G,t)$ of the Madelung equations, when the Planck constant $\hbar$ is made to tend to 0. We will restrict ourselves to \textit{\textbf{``prepared non-discerned quantum systems''}}.

\begin{Definition} - A quantum system, subjected to a potential $V(\mathbf{x})$, is said to be a \textbf{prepared non-discerned} quantum systems, if we know at the initial time, only the density of the initial probability
$\rho^{\hbar}_{0}(\mathbf{x}_G)$ and the initial action $S^{\hbar}_{0}(\mathbf{x}_G)$ of its external wave function, and that these are functions $\rho_{0}(\mathbf{x}_G)$ and $S_{0}(\mathbf{x}_G)$ which are independent of $\hbar$.
\end{Definition}

It is the case of a set of particles without interaction between each other and prepared in the same way: sources of free particles or in a linear field as in the Shimizu~\cite{Shimizu1992} experiment with cold atoms, sources of fullerenes, neutrons, electrons or $C_{60} $ in a
Young's slit experiment. One can then demonstrate~\cite{Gondran2021} theorem below:

\begin{Theoreme}\cite{Gondran2021} \label{r-th1}- When $\hbar$ tends to 0, 
if the external wave function is prepared non-discerned, 
the density
$\rho^{\hbar}(\mathbf{x}_G,t)$ and the action $S^{\hbar}(\mathbf{x}_G,t)$, solutions to Madelung equations
(\ref{eq:Madelung1}-\ref{eq:Madelung3}),
converge to $
\rho(\mathbf{x}_G,t)$ and $
S(\mathbf{x}_G,t)$,
solutions to Hamilton-Jacobi statistical equations:
\begin{eqnarray}\label{eq:statHJ1b}
\frac{\partial S\left(\mathbf{x}_G,t\right) }{\partial
t}+\frac{(\nabla S(\mathbf{x}_G,t) )^{2}}{2m}+V(\mathbf{x}_G)&=&0
\\
\label{eq:statHJ2b}
S(\mathbf{x}_G,0)&=&S_{0}(\mathbf{x}_G)
\\
\label{eq:statHJ3b}
\frac{\partial \mathcal{\rho }\left(\mathbf{x}_G,t\right) }{\partial
t}+ div \left( \rho \left( \mathbf{x}_G,t\right) \frac{\nabla
S\left( \mathbf{x}_G,t\right) }{m}\right)&=&0
\\
\label{eq:statHJ4b}
\rho(\mathbf{x}_G,0)&=&\rho_{0}(\mathbf{x}_G) 
\end{eqnarray}
\end{Theoreme}

\subsection{Action-particle duality in classical mechanics}

This Hamilton-Jacobi action $S(\mathbf{x},t) $ is written from the Hamilton-Jacobi action at the initial time $S_0(\mathbf{x})$ and from the Euler-Lagrange actions, $S_{EL}(\mathbf{x}_0; \mathbf{x},t))$, between all the possible paths from ($\mathbf{x}_0$,0) to ($\mathbf{x}$,t)
\begin{equation}\label{eq:hj}
S(\mathbf{x},t)=min_{\mathbf{x}_0}(S_0(\mathbf{x}_0)+ S_{EL}(\mathbf{x}_0; \mathbf{x},t))
\end{equation}
as a min-plus path integral \cite{Gondran2014c,Kenoufi2020} in a manner \textbf{analogous to Feynman's path integral}, but in the min-plus analysis, a non-linear analysis we have developed \cite{Gondran1996a,Gondran2004a} following Maslov \cite{Maslov1989, Maslov1992}.

These statistical Hamilton-Jacobi equations (equations~(\ref{eq:statHJ1b}-\ref{eq:statHJ4b}) of theorem~\ref{r-th1}) correspond to a set of classical particles, without interaction between each other and subjected to an external potential field $V(\mathbf{x}_G)$, and of which we only know, at the initial time, the probability density $\rho _{0}\left( \mathbf{x}_G\right)$ and
the velocity field $\mathbf{v_{0}(\mathbf{x}_G)}$ through the intermediary
of the initial action $S_0(\mathbf{x}_G)$ (with $\mathbf{v_{0}}(\mathbf{x}_G)=
\frac{\nabla S_0(\mathbf{x}_G)}{m}$). 

These are non-discerned prepared classical particles like the non-discerned prepared quantum particles. For these non-discerned prepared classical particles, the velocity of the center-of-mass of the classical particle is given at each point $ \left(
\mathbf{x}_G, t\right)$ by:
\begin{equation}\label{eq:eqvitesse1}
\mathbf{v}\left( \mathbf{x}_G, t\right) =\frac{\mathbf{\nabla }S\left( \mathbf{%
x}_G, t\right) }{m}\text{\ \ }
\end{equation}

This equation shows that the solution $S\left(
\mathbf{x}_G, t\right) $ to the Hamilton-Jacobi equations (\ref{eq:statHJ1b})
defines the velocity field at any point ($\mathbf{x},t$) from the velocity field $\frac{\nabla S_0(\mathbf{x})}{m} $ at the initial time. Thus, if 
given the initial position $X_G(0)$ of the center-of-mass of an prepared non-discerned classical particle, we deduce from (\ref{eq:eqvitesse1}) the
trajectory $X_G(t)$ of the center-of-mass of the particle by the evolution equation:

\begin{equation}\label{eq:eqvitesse1b}
\dfrac{dX_G(t)}{dt} =\frac{\mathbf{\nabla }S\left( \mathbf{%
x}_G, t\right) }{m}\vert_{\mathbf{x}_G= X_G(t)}
\end{equation}

The Hamilton-Jacobi action $S\left(
\mathbf{x}_G, t\right)$ is thus a field which \textbf{drives the motion of the center-of-mass of the classical particle}.

\subsection{The dBB interpretation of the external wave function}

To interpret the external wave function, we will use the mathematical continuity between the density and action of quantum mechanics and the density and action of classical mechanics, solutions to the Hamilton-Jacobi statistical equations. The \textbf{classical particles} which verify the Hamilton-Jacobi statistical equations have two properties:
\begin{itemize}
    \item They are prepared non-discerned because we do not know at the initial time the position of their centers of mass, but only their initial distribution $\rho_0(\mathbf{x}_G)$. In classical mechanics, we remove the indeterminacy by
adding the initial position of the center-of-mass $X_G(0)$.
    \item They are piloted by the gradient of the Hamilton-Jacobi action, which corresponds to a field that \textit{pilots} the center-of-mass with the equation (\ref{eq:eqvitesse1b}).
\end{itemize}

\textbf{The mathematical continuity} incites a continuity of interpretation by extending these two properties to the centers of mass of quantum particles that verify the Schrödinger equations of the external wave function:
\begin{itemize}
    \item They are prepared non-discerned because we do not know at the initial time the position of their centers of mass, but only their initial distribution $\rho_0(\mathbf{x}_G)$. As in classical mechanics, we remove the indeterminacy by giving the initial position of the center-of-mass $X^h_G(0)$.
    \item The centers of mass of quantum systems are guided by the gradient of the quantum action, which corresponds to a field that \textit{``pilots''} the center-of-mass by the equation:
\begin{equation}\label{eq:vitessequantique}
\textbf{v}^{\hbar}(\mathbf{x}_G,t) = \frac{1}{m} \nabla S^\hbar (\mathbf{x}_G,t)
\end{equation}
\end{itemize}

Thus, the external wave function \textit{pilots} the centers of mass of the quantum systems and corresponds to the \textbf{De Broglie-Bohm interpretation} \cite{deBroglie1927,Bohm1952} restricted to the external wave function only. It is therefore necessary to add the initial position $X^h_G(0)$ of the center-of-mass as well as the equation that gives its evolution:
\begin{equation}\label{eq:vitessequantique2}
\dfrac{dX^h_G(t)}{dt}= \frac{1}{m} \nabla S^\hbar (\mathbf{x}_G,t)\vert_{\mathbf{x}_G= X^h_G(t)}
\end{equation}

When the potential $V(\mathbf{x}_G)$ can be considered as a linear gravitational potential, the Schr\"{o}dinger equations~(\ref{eq:schrodinger1}-\ref{eq:solschrodinger4}) can be solved explicitly and one proves that the dBB trajectories converge to Newton trajectories~\cite{Holland1993, Gondran2021}.
For our approach, the external wave function has the same meaning as the total wave function  for the dBB theory~: the wave function of the particle pilots its center-of-mass. 
Louis de Broglie developed the \textbf{wave pilot theory} in 1926 for the total wave function. He presented it to the Solvey Congress in 1927~\cite{Bacciagaluppi2009} as part of the theory of the double solution he was working on.
However, he abandoned it a few years later. David Bohm rediscovered it in 1952, it is then also known as \textbf{Bohmian mechanics}~\cite{Goldstein}. 
It represents a causal, realistic and observer-free alternative to the Copenhagen interpretation~\cite{Holland1993,Goldstein1998a}.  
Louis de Broglie then resumed his work on this theory, which we call dBB theory. Since 2000, many studies have been carried out to develop and deepen the dBB approach both in theory and in applications.

Numerous simulations of dBB trajectories have been carried out by Sanz and Miret-Artès' teams~\cite{Sanz2002,Sanz2012} or for  quantum chemistry applications~\cite{Wyatt2005}. The development of weak measurements paves the way for an possible experimental validation of these trajectories~\cite{Kocsis2011}. The approaches of Bohmian mechanics to be compatible to the quantum field theory are numerous~\cite{Struyve2010, Durr2004b}. 
A synthesis of the dBB theory can be found in the books of Bricmont~\cite{Bricmont2016,Bricmont2017} or of~\cite{Norsen2017,Oriols2012} and in Goldstein's article~\cite{Goldstein} in the Stanford Encyclopedia of Philosophy.





\section{The internal wave function: the corpuscle of the field-corpuscle duality}
\label{section:InternalWave}

For a quantum system corresponding to an atom or a molecule, the external wave function $\psi(\mathbf{x}_G,t) $ is defined in the laboratory reference frame while the relative wave function $ \varphi(\mathbf{x}'_1,\mathbf{x}'_2,..,\mathbf{x}'_N,t) $ is defined in the center-of-mass reference frame.

\begin{Definition} - From the relative wave function 
$ \varphi(\mathbf{x}'_1,\mathbf{x}'_2,..,\mathbf{x}'_N,t)$ with $N>1$ and the position of the center-of-mass $X^h_G(t)$, we define at each time, $t$, \textbf{the internal wave function }$\Phi(\mathbf{x}_1,\mathbf{x}_2,..,\mathbf{x}_N,t) $ in the laboratory reference frame by the equation:
\begin{equation}\label{eq:foidef}
\Phi(\mathbf{x}_1,\mathbf{x}_2,..,\mathbf{x}_N,t)\equiv \varphi(\mathbf{x}_1 -X^h_G(t),\mathbf{x}_2 -X^h_G(t),..,\mathbf{x}_N - X^h_G(t),t)
\end{equation}
\end{Definition}
We therefore make the following assumption~:
\begin{Hypothese} - \textbf{Hypothesis of the double scale theory} : 
The state of an $N$-body quantum system is simultaneously described by two wave functions in laboratory reference frame~:
\begin{itemize}
    \item the external wave function~: $\psi(\mathbf{x}_G,t)$, a wave related to its center-of-mass~: $X^h_G(t),t)$
    \item the internal wave function : $\Phi(\mathbf{x}_1,\mathbf{x}_2,..,\mathbf{x}_N, t)$.
\end{itemize}
\end{Hypothese}

Let us complete the definition of the internal wave function for a quantum system corresponding to a single particle, $N=1$. This particle admits an external wave function $\psi(\mathbf{x}_G,t)$ with a center-of-mass $X^h_G(t) $ and a relative function $ \varphi(\mathbf{x}',t) $. We define at each time, $t$, its \textbf{internal wave function} $\Phi(\mathbf{x},t) $ in the  laboratory reference frame by the equation:
\begin{equation}\label{eq:foidefpu}
\Phi(\mathbf{x},t)\equiv \varphi(\mathbf{x} -X^h_G(t),t).
\end{equation}

There are therefore two wave functions in the laboratory reference frame, the external wave function and the internal wave function. We have shown that the external wave function is a field that drives the center-of-mass of the corpuscle. Let us now study the convergence of the internal wave function when $\hbar\to0$.

\subsection{Convergence of the internal wave function}

As the size of a single particle tends to zero with $h$, the density $\vert\Phi(\mathbf{x}, t)\vert^2$ of its internal wave function converges mathematically to the Dirac distribution $\delta(\mathbf{x} -X_G(t))$ .   

More generally, if this internal wave function corresponds to an atom or a molecule, its size depends on $h$ and will tend towards zero if we make $ h \rightarrow 0$. We deduce that the density $|\Phi(\mathbf{x}_1,\mathbf{x}_2,..,\mathbf{x}_N, t)|^2$ of the internal wave function converges mathematically to the Dirac distribution at point $\mathbf{x}=X_G(t)$ when we make $ h \rightarrow 0$.
This convergence is completely different from the external wave function

These convergences are postulated and not demonstrated for the general case. However for particular cases, the demonstration is possible, as we recall below for the coherent states of the two dimensional harmonic oscillator. 
For coherent states, let us consider the time dependent Schrödinger equation with the potential $V(\mathbf{x})=\frac{1}{2}m \omega^{2}(x^{2}+y^2)$ and the initial wave function:
\begin{equation}\label{eq:ondeinitiale2d}
\Psi_{0}(\mathbf{x})=\left( 2\pi \sigma _{\hbar}^{2}\right)
^{-\frac{1}{2}}e^{-\frac{ ( \mathbf{x}-\mathbf{x}_{0})
^{2}}{4\sigma _{\hbar}^{2}}+i \frac{m \textbf{v}_0 .
\mathbf{x}}{\hbar} } =
\sqrt{\rho^{\hbar}_{0}(\mathbf{x})}\exp\left(i\frac{S^{\hbar}_{0}(\mathbf{x})}{\hbar}\right)
\end{equation}
where $\rho^{\hbar}_{0}(\mathbf{x})= ( 2\pi \sigma _{\hbar}^{2}) ^{-1}
e^{-\frac{( \mathbf{x}-\mathbf{x}_{0}) ^{2}}{2\sigma
_{\hbar}^{2}}}$ and $S_{0}(\mathbf{x})=S^{\hbar}_{0}(\mathbf{x})= m \textbf{v}_0 \cdot
\mathbf{x}$ are respectively
the initial density and action, 
with $\sigma_\hbar=\sqrt{\frac{\hbar}{2 m \omega}}$ and where $\mathbf{x}_0$ and $\textbf{v}_0$ are the initial position and velocity of the center of the wave packet.

Therefore, the wave function $\Psi (\mathbf{x},t)$ is:
\begin{equation}
\Psi(\mathbf{x},t)=\left( 2\pi \sigma _{\hbar}^{2}\right)
^{-\frac{1}{2}}
  e^{-\frac{( \mathbf{x}- \mathbf{x}(t)) ^{2}}{4\sigma
_{\hbar}^{2}}+i \frac{m \textbf{v}(t). \mathbf{x}
-g^\hbar(t)}{\hbar} }
\end{equation}
where $\mathbf{x}(t)$ and $\textbf{v}(t)$ correspond to the
position and velocity of a classical particle subjected to the
potential $V(\mathbf{x})=\frac{1}{2} m \omega^{2} (x^2 +y^2)$ where  $ g^\hbar(t)=\int _0 ^t ( \hbar \omega +
\frac{1}{2} m \textbf{v}^{2}(s) - \frac{1}{2} m \omega^{2}
\mathbf{x}^{2}(s)) ds$.
We deduce the value of the density
$\rho^{\hbar}(\mathbf{x},t)$ and of the action
$S^{\hbar}(\mathbf{x},t)$:
\begin{equation}\label{eq:densiteaction}
\rho^{\hbar}(\mathbf{x},t)=\left( 2\pi
\sigma_{\hbar} ^{2} \right) ^{-1}e^{- \frac{(
\mathbf{x}-\mathbf{x}(t)) ^{2}}{2\sigma_{\hbar} ^{2} }} \qquad and \qquad S^{\hbar}(\mathbf{x},t)=    m
\textbf{v}(t)\cdot \mathbf{x} -  g^{\hbar}(t).
\end{equation}
We deduce the following theorem by considering the limit $\hbar\to0$ of equation~(\ref{eq:densiteaction}).
\begin{Theoreme}
- \label{t-convergenceparticulediscerne} Let the coherent states of the two-dimensional harmonic oscillator defined by equations (\ref{eq:ondeinitiale2d}).
When $\hbar$ tends to 0, then $\sigma_{\hbar}$ tends to
0, and the density
$\rho^{\hbar}(\mathbf{x},t)$ and the action
$S^{\hbar}(\mathbf{x},t)$,
converge respectively to
\begin{equation}\label{eq:delta}
\rho(\mathbf{x},t)=\delta( \mathbf{x}-
\mathbf{x}(t)) \qquad and \qquad
S(\mathbf{x},t)= m \textbf{v}(t)\cdot \mathbf{x} -g(t)
\end{equation}
with $g(t)=\int _0 ^t ( \frac{1}{2} m \textbf{v}^{2}(s) -
\frac{1}{2} m \omega^{2} \mathbf{x}^{2}(s)) ds$ and where  $S(\mathbf{x},t)$ is the Hamilton-Jacobi singular action.
\end{Theoreme}

The wave function therefore converges to the motion
$\mathbf{x}(t)$ of a single classical \textbf{oscillator}. This
classical particle is \textbf{discerned} because it is completely defined by its initial condition
initial condition $\mathbf{x}_0$ and its initial action
$S_0(\mathbf{x}) = m \textbf{v}_0 . \mathbf{x}$.

Note that one should speak of \textit{the center-of-mass of the classical particle} rather than of \textit{the classical particle}. In this framework, equation (\ref{eq:delta}) corresponds to the center-of-mass equation and 
is written~:
\begin{equation}\label{eq:deltab}
\rho(\mathbf{x}_G,t)=\delta( \mathbf{x}_G-
X_G(t)) \qquad and \qquad
S(\mathbf{x}_G,t)= m X_G'(t)\cdot \mathbf{x}_G -g(t)
\end{equation}
with initial position $X_G(0)=\mathbf{x}_0$ and initial velocity $X_G'(0)= \textbf{v}_0$.

\subsection{Interpretation of the internal wave function for an elementary particle}

The interpretation that we propose for the internal wave function is the one proposed by Schrödinger in 1926~\cite{Schrodinger1926} and at the Solvay congress of 1927~\cite{Schrodinger1927}. It is based on the solution of the coherent states of the harmonic oscillator, which is consistent with the second quantization and with the introduction of the creation and annihilation operators. These coherent states of the
harmonic oscillator are very particular, as Schrödinger pointed out~\cite{Schrodinger1926}:
\textit{``Our wave packet always remains grouped, and does not spread over an increasingly large space over time, as do, for example, wave packets that we are used to in optics.''}
 This wave packet corresponds to a single particle (soliton) which keeps its shape and whose center-of-mass follows a periodic trajectory identical to that of the center-of-mass of a classical harmonic oscillator. Schrödinger thinks that this interpretation extends to the electron of the hydrogen atom \cite{Schrodinger1926}:
\textit{``It is certain that it is possible to construct by a process quite similar to the previous one, wave packets gravitating on Kepler ellipses at a large number of quanta and forming the wave image of the electron of a hydrogen atom; but in this case the difficulties of calculation will be much greater than in the particularly simple example that we have treated here and which from this point of view is almost a classroom exercise''.}

\begin{Hypothese}- \textbf{Schrödinger conjecture for an elementary particle}: An elementary particle, such as a free electron or a bound electron in a hydrogen atom, can be considered as an extended particle whose density is given by the square $\mid \Phi(\mathbf{x},t) \mid^2$ of the internal wave function in the laboratory reference frame.
\end{Hypothese}
The position $X_G(t)$ of the center-of-mass of the particle is then obtained from the internal wave function by:
\begin{equation}\label{eq:cmpar}
X_G(t)=\int \mathbf{x} |\Phi(\mathbf{x},t)|^2 d\mathbf{x}
\end{equation}

\textbf{The simultaneous existence of the two wave functions makes quantum mechanics complete.} Indeed, it is not necessary to add the initial position of the center-of-mass $X_G(0)$ as in the de Broglie-Bohm interpretation of the external wave function because it is defined by the initial internal wave function: $ X_G(0)=\int \mathbf{x} \vert\Phi_0(\mathbf{x})|^2 d\mathbf{x}$. 

The dBB theory can be seen as a specific case of the double-scale theory if the internal function is reduced to a point, the center-of-mass of the particle; i.e., $\Phi(\mathbf{x})$ is a Dirac function~: $\Phi(\mathbf{x},t)=\delta(\mathbf{x}-X_G(t), t)$.

\begin{Remarque}\label{rem:cm2}
The change of variable $\Phi(\mathbf{x},t)\equiv \varphi(\mathbf{x} -X^h_G(t),t)$
is very important. Indeed, if the average is compute with $\varphi(\mathbf{x},t)$, then, by definition, we obtain~:
\begin{equation}\label{eq:cmpar2}
\int \mathbf{x} |\varphi(\mathbf{x},t)|^2 d\mathbf{x}=0
\end{equation}
\end{Remarque}



\begin{Remarque}\label{rem:etendu} - The hypothesis of an extended particle obliges to introduce forces to maintain the cohesion of the particle. A model of the extended and deformable electron is proposed by Poincaré in his famous Palermo memoir \cite{Poincare1906}.  Dirac makes the same argument in his article of 1962 \textit{``An extensible model of the electron''}~\cite{Dirac1962}. It is also the basis of Delmelt's work on the size of the electron~\cite{Dehmelt1989}.
\end{Remarque}
 The spatial extension of the electron is also consistent with the introduction of cut-offs in quantum electrodynamics (QED) to eliminate the infinities due to the hypothesis of point particles.

\subsection{Interpretation of the internal wave function for a system of $N$ particles}

At the Solvay Congress in 1927, Schrödinger generalized his interpretation for the wave function of a single particle to that of a wave function of $N$ particles in configuration space:

\begin{quotation}
\textit{``I found the following way of looking at things useful; it may be a little naïve but it is easy to grasp.
The classical system of material points does not really exist, but there is something that continuously fills all the space [...] the real system is a composite image of the classical system in all its possible states, obtained by using $\phi \phi^*$ as a ``weight function''.
The systems to which the theory is applied are classically composed of a large number of charged material points. As we have just seen, the charge of each of these points is distributed continuously through space and each charge point e provides the contribution of the $e \int \phi  \phi^* dx dy dz$ to the charge of the quarterly volume element $dx dy dz$. 
As $\phi  \phi^*$ generally depends on time, these charges vary.''}
\end{quotation}
Schrödinger took up this interpretation in 1952~\cite{Schrodinger1952} and was strongly criticized by the Copenhagen school and in particular by Born in his 1954 Nobel lecture \cite{Born1955}:
\begin{quotation}
\textit{``Schrödinger thought that his wave theory made it possible to return to deterministic classical physics. He proposed (and he has recently emphasized his proposal anew), to dispense with the particle representation entirely, and instead of speaking of electrons as particles, to consider them as a continuous density distribution $|\psi|^2 $ (or electric density $e|\psi|^2$).''}
\end{quotation}

The criticism related to the contradiction between Schrödinger's definition of the particles as \textit{``narrow wave packets''} and the fact that the external function spreads out in time. One can understand the criticism of such an interpretation if one does not differentiate between the internal and external wave functions; but this argument is no longer valid for the internal wave function. 

We extend Schrödinger's approach by postulating that the internal wave function of an $N$-body system is the product of $N$ \textit{narrow wave packets} each corresponding to a body of the system. We make the following assumption.

\begin{Hypothese}\label{hypo:GeneralizedSchrodinger}- \textbf{Generalized Schrödinger conjecture for a system of N particles} - To each internal wave function in the laboratory reference frame $ \Phi (\mathbf{x}_1,\mathbf{x}_2,\mathbf{x}_3,... ,\mathbf{x}_N,t) $ of a quantum system composed of $N$ particles, one can associate $N$ individual time-dependent wave functions in the laboratory reference frame $\Phi^j(\mathbf{x},t) $ such that $ \mid \Phi^j(\mathbf{x},t)\mid^2 $ represents the density of the particle $j=1..N$. The second part of the Hypothesis is to assume the following equation:
\begin{equation}\label{eq:foiNa}
\Phi(\mathbf{x}_1,\mathbf{x}_2,\mathbf{x}_3,...,\mathbf{x}_N,t)= \prod_{j=1}^{N} \Phi^j(\mathbf{x}_j,t), 
\end{equation}
\end{Hypothese} 

This last assumption consider that the extended particles, i.e. the individual wave functions $\Phi^i(\mathbf{x},t)$, do not overlap in real space and therefore  have disjoint supports~: $\Phi^j(\mathbf{x},t)\Phi^i(\mathbf{x},t)=0 $ for all $i\neq j$.  This is a property that is also verified by the individual relative Schrödinger wave functions $\varphi^j(\mathbf{x},t)= \Phi^j(\mathbf{x} +X^h_G(t),t) $. 
\begin{Remarque} - With this assumption, the internal wave function $ \Phi(\mathbf{x}_1,\mathbf{x}_2,\mathbf{x}_3,...,\mathbf{x}_N, t) $ in the configuration space is replaced by N individual (internal) wave functions in the real three dimensional space.
\end{Remarque}

\begin{Remarque}- 
Hypothesis~\ref{hypo:GeneralizedSchrodinger} is for example valid for the nuclei of the atoms of a molecule.
This assumption is not respected for the stationary wave functions of the electrons of an atom because the orbitals can overlap.
Nevertheless, we think that this assumption is valid if the electrons wave functions are time-dependent internal wave functions.

\end{Remarque}

We complete the hypothesis~\ref{hypo:GeneralizedSchrodinger} with the following hypothesis:

\begin{Hypothese}- We assume that the N individual time-dependent wave functions $\varphi^j (\mathbf{x}_j,t)$ are solutions to the N individual non linear Schrödinger equations:
\begin{equation}\label{eq:schrodinger2bc34}
i\hslash \frac{\partial \varphi^j(\mathbf{x}_j,t) }{\partial t}=- \frac{\hbar ^{2}}{2 m_j}
\Delta_{\mathbf{x}_j} \varphi^j(\mathbf{x}_j,t) + \left(\sum_{\underset{i \neq j}{i=1}}^N
 \int d\mathbf{x}_i \vert  \varphi^i(\mathbf{x}_i,t)  \vert^2 U_{ji}(\vert \mathbf{x}_j -\mathbf{x}_i \vert)\right) \varphi^j(\mathbf{x}_j,t)
\end{equation}
with the initial conditions:
\begin{equation}\label{eq:solschrodingervarphi}
 \varphi^j(\mathbf{x}_j,0)=  \varphi^j_0(\mathbf{x}_j)= \Phi^j_0(\mathbf{x}_j+X^h_G(0)).
\end{equation}
\end{Hypothese}-

Because the size of each particle $i=1...N$ (i.e. the support of $\varphi^i$) is very small compared to the size of a atom or molecule, we can assume that we have for all $t$, 
$\vert \Phi^i(\mathbf{x},t) \vert^2 \simeq \delta (\mathbf{x} - X^i(t))$ and $\vert \varphi^i(\mathbf{x},t) \vert^2 \simeq \delta (\mathbf{x} - \mathbf{x}^i(t))$ where $\mathbf{x}^i(t)= \int \textbf{x } |\varphi^i(\mathbf{x},t)|^2 d\mathbf{x}$ is the position of the center-of-mass of particle $i$ in the center-of-mass reference frame and $X^i(t)= \int \textbf{x } |\Phi^i(\mathbf{x},t)|^2 d\mathbf{x} $ in the laboratory reference frame with $ X^i(t)= \mathbf{x}^i(t) +X^h_G(t)$. The integral of equation~(\ref{eq:schrodinger2bc34}) can then be approximated by: $\int d\mathbf{x}_i \vert  \varphi^i(\mathbf{x}_i,t)  \vert^2 U_{ji}(\mathbf{x}_j -\mathbf{x}_i)\approx U_{ji}(\vert \mathbf{x}_j -\mathbf{x}^i(t) \vert)$ and we obtain the following system of equations:

\begin{equation}\label{eq:schrodinger2bc36}
i\hslash \frac{\partial \varphi^j(\mathbf{x}_j,t) }{\partial t}=- \frac{\hbar ^{2}}{2 m_j}
\Delta_{\mathbf{x}_j} \varphi^j(\mathbf{x}_j,t) + \left(\sum_{\underset{i \neq j}{i=1}}^N
 U_{ji}(\vert \mathbf{x}_j -\mathbf{x}^i(t) \vert)\right) \varphi^j(\mathbf{x}_j,t)
\end{equation}
\begin{equation}\label{eq:schrodinger2appro}
\mathbf{x}^i(t)= \int \mathbf{x} \vert \varphi^i(\mathbf{x},t) \vert^2d\mathbf{x}
\end{equation}
with the initial conditions~:
\begin{equation}\label{eq:solschrodinger5}
 \varphi^j(\mathbf{x}_j,0)=  \varphi^j_0(\mathbf{x}_j)
\end{equation}

The approximate calculation of equations (\ref{eq:schrodinger2bc36}-\ref{eq:solschrodinger5}) can be obtained from Ehrenfest's theorem.

There are many arguments, both experimental and theoretical, in favor of this generalized Schrödinger interpretation for the internal wave function (Hypothesis~\ref{hypo:GeneralizedSchrodinger}):
\begin{itemize}
    \item It is fundamentally compatible with the methodology of \textbf{molecular dynamics}. Indeed, if we separate the nuclei from the electrons in a molecule, the nuclei will have internal wave functions verifying the two Schrödinger conjectures.
    \item It is also fundamentally compatible with the first Hohenberg-Kohn~theorem\cite{Hohenberg1964}, which is the basis of \textbf{density functional theory}, which states that a given electron density corresponds to a unique wave function.
    \item It is also in agreement with the recent (2019) experiment by Minev, Devoret et al.~\cite{Minev2019} on \textit{``the jump from ground state to an excited state of a three-level superconducting artificial atom''}. This experiment seems to prove Schrödinger right in his 1952 discussion with the Copenhagen school on quantum jumps, given that \textit{``the experimental results show that the evolution of each jump made is continuous, coherent and deterministic''}. As Devoret explains: \textit{``Our experimental results show that quantum jumps are unpredictable and discrete (as Bohr thought) over long periods of time, they can be continuous (as suggested by Schrödinger) and predictable over short periods of time''}.
\end{itemize}

\section{A new reading of quantum mechanics}
\label{section:NewReading}

This scale-dependent dual interpretation is a framework for reading quantum mechanics in a simple way and also explains the choices of other interpretations that did not distinguish between external and internal wave functions.

\textbf{The wave-corpuscle duality} corresponds to the simultaneous existence of two wave functions linked together, at each instant $t$, by the position of the center-of-mass $X^h_G(t)$: while the external wave function corresponds to the wave (field) and the internal wave function corresponds to the corpuscle. \textbf{Theoretically}, we therefore have field AND corpuscle. \textbf{Experimentally}, a coherent source of particles is a set of particles which all have the same external wave function, but whose internal wave functions are different.
Thus in the slit diffraction, double slit interference, Stern and Gerlach spin measurement, tunneling effect and EPR-B experiments, the preparation of the quantum system is represented by the external wave function alone. 
The internal wave function, which is unknown, explain the statistical results of Born's interpretation.


\textbf{Quantum mechanics is complete} if the state of a quantum system corresponds to the simultaneous existence of both wave functions, incomplete otherwise. The position of the center-of-mass in the external wave function that must be added in the de Broglie-Bohm interpretation is obtained from the internal wave function by the equations (\ref{eq:cmpar}) and (\ref{eq:schrodinger2appro}). We can thus interpret Everett's multiple worlds as the set of internal wave functions compatible with the external wave function. 

  

\textbf{The measurement corresponds to the impact of the internal wave} on the detection screen. It is thus the internal wave function that is involved in the collapse of the wave packet. 
The Copenhagen interpretation, which does not differentiate between the external and internal waves, cannot therefore see that the reduction of the wave packet concerns mainly the internal wave function. For the same reason, the GRW interpretation~\cite{Ghirardi1986} requires an objective non-linear perturbation. This perturbation exists because the particle is stopped by the detection screen following a non-linear absorption phenomenon. In the double slit experiment, \textbf{the external wave function} passes \textbf{through both slits} while \textbf{the internal wave function} passes through \textbf{only one slit}.

\textbf{Heisenberg's inequalities} correspond, for the external wave function, to \textbf{uncertainty relations} on the positions and velocities of the centers of mass of an ensemble of molecules admitting this same wave function. 
For the internal wave function, the Heisenberg equalities correspond to \textbf{indeterminacy relations} on the different positions and velocities of this extended particle \cite{Gondran2021}.

\textbf{The non-local hidden variable} of Bell's theorem and the EPR-B experiment concerns the external wave function \cite{Gondran2021} while the position of the impacts is a \textbf{local measured variable} corresponding to the internal wave function.

This scale-dependent dual solution can be tested experimentally by an asymmetric double slit experiment such as those we proposed a few years ago \cite{Gondran2008b} and are currently pursuing \cite{Gondran2023a}.

\section{Conclusion}

We have proposed an experimentally testable interpretation of non-relativistic quantum mechanics that gives a new understanding of the links between quantum and classical mechanics.
This scale-dependent double solution seems strongly consistent with the specifications of Louis de Broglie who wrote in 1971:
\begin{quotation}
\textit{``I introduced,
under the name of "theory of the double solution", the idea
that it was necessary to differentiate between two solutions, distinct but
intimately connected to the wave equation, one of which I called the $u$-wave, being a real and non-normalizable physical wave
with a local accident defining the particle and represented by a singularity, the other one, Schrödinger's $\psi$ wave, normable and without singularity, which would be
only a representation of probabilities.''} \cite{deBroglie1971}
\end{quotation}
Our external wave function corresponds to the Schrödinger's $\psi$ wave and our internal wave function corresponds to the u-wave.


We believe that this new reading grid allows us to extend this realistic interpretation to all quantum mechanics as well as to relativity. We are indeed preparing an extension of this double scale theory to the relativistic case by considering the Gordon decomposition of the Dirac equation as a convection current corresponding to the external wave function (large components) and a spin current corresponding to the internal wave function (small components)~\cite{Gondran2023b}. We will also propose a semi-classical gravity converging to the Newton gravity when $\hbar\to0$~\cite{Gondran2022a}.
An extension to the second quantization~\cite{Gondran2023c}
also seems possible by considering the external wave function as a field and the internal wave function as extended particles that can be created or annihilated by being bound to the excitations of this field~\cite{Gondran2023c}.

This interpretation of quantum mechanics may also be considered as an answer to one of the Einstein's final texts on the interpretation of quantum mechanics 
(1953), \textit{Elementary Considerations on the Interpretation of the Foundations of Quantum Mechanics}
in homage to Max Born~\cite{Einstein1953}:
\begin{quotation}
\textit{``\textbf{The first effort} goes back to de Broglie and has been pursued further by Bohm with great perspicacity [...] \textbf{The second attempt}, which aims at achieving a "real description" of an individual system, based on the Schrödinger equation, has been made by Schrödinger himself. Briefly, his ideas are as follows. \textbf{The $\psi$-function itself represents reality,} and does not stand in need of the Born interpretation[...] From the previous considerations, it follows that the only acceptable interpretation of Schrödinger's equation up to now is the statistical interpretation given by Born. However, it does not give the "real description" of the individual system, but only statistical statements related to sets of systems.''}
\end{quotation}

\bibliographystyle{elsarticle-num}
\bibliography{biblio_mq}

\end{document}